\documentclass[twoside]{LCWS11}
\usepackage[latin1]{inputenc}
\usepackage[dvips]{graphicx,epsfig,color}
\usepackage{wrapfig,rotating}
\usepackage{amssymb,amsmath,array}
\usepackage{cite}
\begin{document}
\title{A RADIATIVELY IMPROVED FERMIOPHOBIC \\
\vspace{0.2cm}
HIGGS BOSON SCENARIO
%%%%   Paper title goes here  %%%%%%%%%%%%%%
} %% 
%***********************************************************************
% AUTHORS INFORMATION AREA
%***********************************************************************
\author{E. Gabrielli$^1$ and B. Mele$^2$
% Optional short acknowledgment: remove next line if non-needed
\thanks{This work was supported by the ESF 
grant MTT60 and by the recurrent financing SF0690030s09 project.}
% DO NOT MODIFY THE FOLLOWING '\vspace' ARGUMENT
\vspace{.3cm}\\
% Addresses and institutions (remove "1- " in case of a single institution)
1- National Institute of Chemical Physics and Biophysics, \\
Ravala 10, 10143 Tallinn, Estonia\\
%% Remove the next three lines in case of a single institution
\vspace{.1cm}\\
2- INFN, Sezione di Roma, c/o Dipartimento di Fisica,\\
 Universit\`a di Roma  ``La Sapienza'', P. le A. Moro 2, I-00185 Rome, Italy
}
%%***********************************************************************
% END OF AUTHORS INFORMATION AREA
%***********************************************************************

\maketitle

\begin{abstract}
The naive fermiophobic scenario is unstable under radiative corrections, due to
the chiral-symmetry breaking induced by fermion mass terms.
In a recent study, the problem of including the radiative corrections has been 
tackled via an effective field theory approach. The renormalized 
Yukawa couplings are assumed to vanish at a high energy scale
$\Lambda$, and their values at the electroweak scale are computed via modified  
Renormalization Group Equations. We show  that, 
in case a fermiophobic Higgs scenario shows up at the LHC, a 
linear collider program will be needed to accurately measure
the radiative Yukawa structure, and consequently constrain the $\Lambda$ scale.
\end{abstract}

\section{Introduction}

The standard model (SM) Higgs boson search \cite{djouadi} is approaching its completion 
with the operation of the LHC at a larger-than-nominal luminosity. 
In a few months,  the ATLAS and CMS experiments will be able either 
to find out a SM-Higgs-like signal (and subsequently start the 
exploration of its properties) or to exclude the existence of a 
SM Higgs boson. In the latter case, exploring the possibility of 
a Higgs boson with characteristics different from the SM ones 
will in general take longer. An exception to this expectation 
could be given by a light fermiophobic  Higgs boson.

A fermiophobic (FP) Higgs boson by definition  is decoupled from fermions. 
As a consequence, its branching ratios (BR's) for decays into photons and 
vector bosons are enhanced for light Higgs masses, where the SM decays into 
fermion pairs are dominant.
In particular, assuming vanishing Higgs widths into fermions in the SM 
framework gives rise to an enhancement of the BR into photon pairs 
by a factor about [110, 30, 10, 5] for the Higgs mass 
$m_H\simeq [100, 110, 120, 130]$ GeV, respectively, 
and would increase all the decay rates into vector bosons 
for $m_H\le 150$ GeV \cite{LHCHiggsCrossSectionWG}. On the other hand, 
the main Higgs production mechanism through gluon fusion, which depends on the Higgs coupling to heavy colored fermions, is missing for the FP Higgs boson, and total production rates are in general depleted. The combination of an enhanced decay rate and a depleted production would anyway make the detection of a FP Higgs lighter than about 120 GeV
remarkably easier than in the SM picture, where the discovery of a Higgs boson
through the depleted $H\to\gamma\gamma$ signature is quite challenging.

\noindent
In general, assuming vanishing Higgs couplings to fermions implies 
a mechanism for the generation of fermion masses different from the SM one. 
In the SM, fermion masses,  as well as  vector boson masses, 
arise as the result of the spontaneous Electroweak-Symmetry Breaking (EWSB)
induced by the Higgs field.
The vacuum expectation value of the Higgs field sets, 
through the Yukawa coupling constants, also the fermion masses, 
and produces spontaneous Chiral Symmetry Breaking (ChSB) 
in the Lagrangian. Decoupling the Higgs boson from fermions 
by switching off Yukawa couplings is presently phenomenological 
acceptable \cite{noi-uno},  but makes the theory 
non-renormalizable.
This implies that in order to compute radiative corrections to 
the "naive" (i.e. tree-level) FP scenario one needs further assumptions.

In a recent paper, we addressed the issue of evaluating the radiative 
Yukawa couplings  in a general FP Higgs scenario  \cite{noi-uno}. 
There are at least two possible ways to proceed.
First, one can build up a renormalizable extension of the SM predicting a 
FP Higgs boson at tree-level. This implies in general an increase 
in the number of the degrees of freedom of the Higgs sector  
\cite{HiggsFP}. Then, radiative 
corrections are computable, although  known models do 
not predict large effects
in the Yukawa couplings since the ChSB scale is  of the order of  
the EWSB scale.
Second, one can try to  find a general description of  a wide class of 
possibilities where  the Higgs mechanism
generates the vector boson masses just as in the SM, 
but where fermion masses have a different (unknown) origin. 
This can be pursued by considering the SM with non-vanishing fermion 
masses but  no Yukawa couplings as an effective field theory
(EFT).

In  \cite{noi-uno}, we adopted the latter (more general) approach
by switching off tree-level Yukawa couplings in the SM Lagrangian.
In particular, while in the SM the ChSB, associated to the 
non-vanishing
 fermion masses, and the  EWSB, associated to the non-vanishing 
vector-boson masses, are both set at the same energy scale by the vacuum
expectation value $v \simeq 246$ GeV of the Higgs field, we
assume  ChSB and fermion masses are
generated by some new  mechanism  at
an energy scale $\Lambda$, that will be in general larger or 
much larger than the  electroweak scale (contributions 
to EWSB arising from the ChSB mechanism are assumed to be small).
Then, we  assume that  $m_H \le  v$ and, for simplicity, only SM degrees 
of freedom propagate at energies below $\Lambda$.
In this framework, the condition of vanishing Yukawa couplings at 
some  energy scale is unstable under radiative corrections. 
Because of ChSB, radiative effects regenerate Yukawa couplings, 
with a control parameter given by fermion masses.

\noindent
We then impose  that 
the Yukawa couplings vanish at the (renormalization) scale
$\Lambda \gg m_H$, which is the renormalization condition  for  
the fermion-Higgs boson decoupling.
In perturbation theory, this condition should be replaced by the finite 
one-loop threshold conditions of the Yukawa couplings at the $\Lambda$ scale, 
calculated in the framework of a renormalizable 
UV completion of the theory above the scale 
$\Lambda$. Since $ m_f \ll \Lambda$,  
$Y_f(\Lambda)$ are expected  (by ChSB and dimensional analysis arguments)  
to be of order ${\cal O}(m_f/\Lambda) \ll 1$,
and $Y_f(\Lambda)=0$ should be seen as a particular approximation. 
Then, we use the EFT approach to compute the universal 
(i.e., independent of the 
particular UV completion of the theory above $\Lambda$)
leading-log components of  the Yukawa couplings
at the $m_H$ scale through Renormalization Group (RG) equations.
The RG equations as computed in the SM \cite{arason} (where $Y_f$'s and $m_f$'s are related by spontaneous ChSB 
induced by the SM Higgs field) do not fit the present framework, 
and new RG equations were derived in \cite{noi-uno} by keeping $Y_f$'s and 
$m_f$'s as independent parameters. The numerical values for the  radiative Yukawa couplings derived in this approach (as well as the results shown below on BR's and cross sections) can be found in \cite{noi-uno,noi-due}.

In Figure~1 (upper-left), the total Higgs width is shown versus $m_H$ for the improved FP-Higgs scenario at different values of the $\Lambda$ scale. The small Higgs total width considerable enhances BR's for radiative decays into fermion pairs, especially at large $\Lambda$ [Figure~1 (upper-right)]. In particular,  BR$(b\bar b)>10\%$ for $\Lambda> 10^7$ GeV.
Also shown, in the same plot, is the BR for the flavor-changing
decay $H\to bs$ as computed in \cite{noi-due}.
In Figure~1 (lower-left), one can see that, for $\Lambda$ near the GUT scale, BR$(b\bar b)$ can almost reach the SM value for  $m_H<110$
GeV.

Present experimental bounds on the improved FP scenario are expected to depend non-trivially on the $\Lambda$ scale assumed.
In particular, at  $\Lambda\sim m_H$ (where log effects vanish in the RG equations) one should recover the $m_H$ bounds
found, assuming a tree-level FP scenario, at LEP \cite{FP-LEP}, Tevatron \cite{FP-Tevatron}, and LHC \cite{FP-CMS}, while at  $\Lambda \sim M_{GUT}$ one should approach the SM Higgs mass limits \cite{SM-LEP,SM-Tevatron,SM-ATLAS,SM-CMS}.
At intermediate $\Lambda$'s, milder bounds are  expected due to the interplay of an enhanced BR($\gamma\gamma$) (that is anyway lower than in the tree-level FP scenario) and a BR($b \bar b$) that is non-vanishing, but less than its SM value.
Just as an example of the $m_H$ bounds behavior versus $\Lambda$, we present in Figure~1 (lower-right), what is obtained by the help of  the code  Higgsbounds  \cite{Higgsbounds}, version 3.4.0beta, after modeling the Higgs couplings via the effective Yukawa couplings
of the  improved FP scenario.

In Figure~2, we present production rates corresponding to different Higgs decay signatures at the LHC at 7 TeV of $pp$ c.m. collision energy (VBF stands for vector-boson-fusion production, while WH refers to the Higgs associated production with a $W$ boson). Dashed (blue) lines show, for reference, the corresponding SM rates,
where the $gg$ fusion gives the dominant mechanism.
Dash-dotted (black) lines refer to the tree-level FP scenario. 
Note that, for a light Higgs, the FP scenario rates overshoot the SM ones, while, for $m_H>130$ GeV, they are considerable lower than
them. This makes the FP scenario evade most of the present Tevatron and LHC bounds at both intermediate and large $m_H$ \cite{SM-Tevatron,SM-ATLAS,SM-CMS}.

In case a FP Higgs  scenario shows up at the LHC with $m_H<150$ GeV, a linear collider program \cite{ILC} would be crucial to achieve the accuracy on the Higgs BR's measurements that are needed to get a precise $\Lambda$ determination \cite{noi-due}.  In Figure~3, we show the production rates corresponding to  Higgs decays into $\gamma\gamma$ (upper-left) and 
$b \bar b$ (upper-right) at an $e^+e^-$ collider with $\sqrt S \simeq 350$ GeV. Note that the sensitivity to $\Lambda$ of the 
$\gamma\gamma$ signature drops quite rapidly when $m_H$ increases. On the other hand,
the $b \bar b$ signature is particularly sensitive to the scale $\Lambda$, and would allow a $\Lambda$ determination up to relatively large Higgs masses (i.e. $m_H\sim 150$ GeV).

A linear collider program would also allow to test the correlation of  BR' s for different fermionic channels.
 Figure~3   
shows the correlations of the Higgs branching ratios 
\{BR($H\to cc$), BR($H\to \tau \tau$)\} (lower-left)
and the flavor-changing  
BR$(H\to bs)$  (lower-right), versus the BR($H\to bb$), for different  values
of the Higgs mass $m_H=110,120,130$ GeV and  $\Lambda$ scale. Flavor changing decay BR's for $H\to bs$ have been computed through RG equations for the off-diagonal Yukawa couplings in the flavor space \cite{noi-due}.
BR$(H\to bs)$ of the order of $(10^{-4}-10^{-3})$, for 
$\Lambda=(10^4-10^{16})$ GeV, have been obtained to be confronted with the corresponding BR$_{SM}(H\to bs)<10^{-7}$.

In conclusion, in the study presented here we stressed that a 
model with a Higgs boson that is decoupled from fermions at tree-level  leads to a framework that is unstable under radiative corrections. Although Yukawa couplings are set to zero at tree level, the non-vanishing fermion masses give rise to ChSB in the Lagrangian that radiatively regenerates Yukawa couplings. We adopted an effective field theory approach to compute radiative corrections in the FP Higgs framework. This might provide a unified description of a wide class of possibilities
for the unknown mechanism of fermion mass generation.
In case the scale $\Lambda$ of ChSB is much larger that the electroweak scale,
the radiatively generated BR($H\to bb$) could approach its SM value. LHC is about to test the present scenario.
In case a FP Higgs pattern shows up with $m_H<150$ GeV, a linear collider will be
crucial to accurately measure BR($H\to bb, cc, \tau\tau$), and determine the $\Lambda$ scale. In particular, the rates
of $e^+e^-\to ZH \to Z b\bar b$ are remarkably sensitive
to $\Lambda$. An analysis of the accuracy expected at a linear collider in the improved FP Higgs scenario would be crucial to set the machine potential to carry out a  measurement of 
the  ChSB scale.

%%%%%%%%%%%%%%%%%%%%%%%%%%%%%%%  FIG 1 %%%%%%%%%%%%%%%%%%%%%%%%%%%%%%%%%%%%%%
\begin{figure}[tpb]
\begin{center}
\centerline{
\epsfxsize=3.1in\epsfig{file=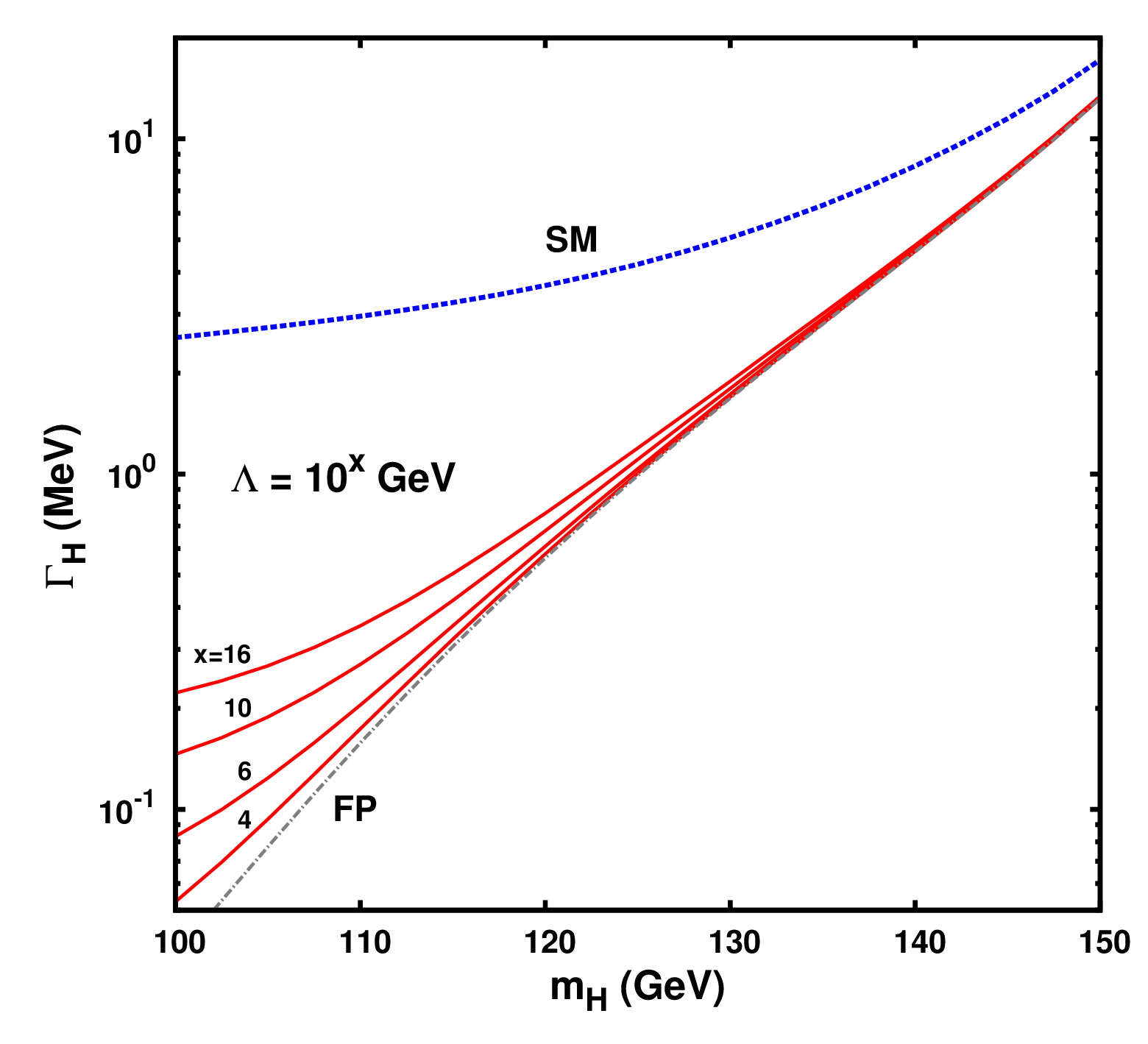, width=6cm,height=7.5cm, angle=-90}
\hspace{0cm}
\hfil
\epsfxsize=3.1in\epsfig{file=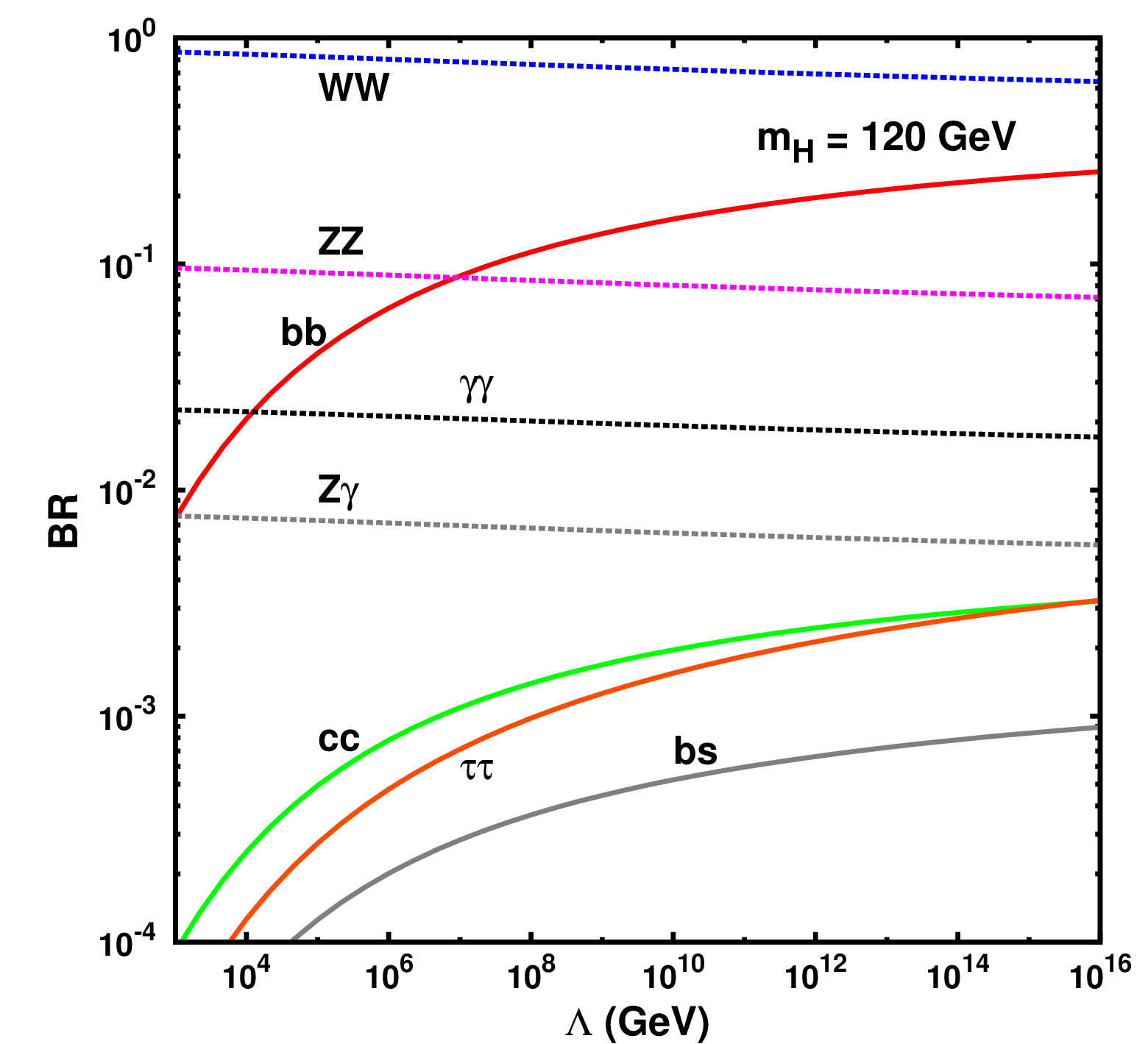,  
width=6cm, height=7.5cm, angle=-90}}
\vspace{0.5cm}
\centerline{
\epsfxsize=3.1in\epsfig{file=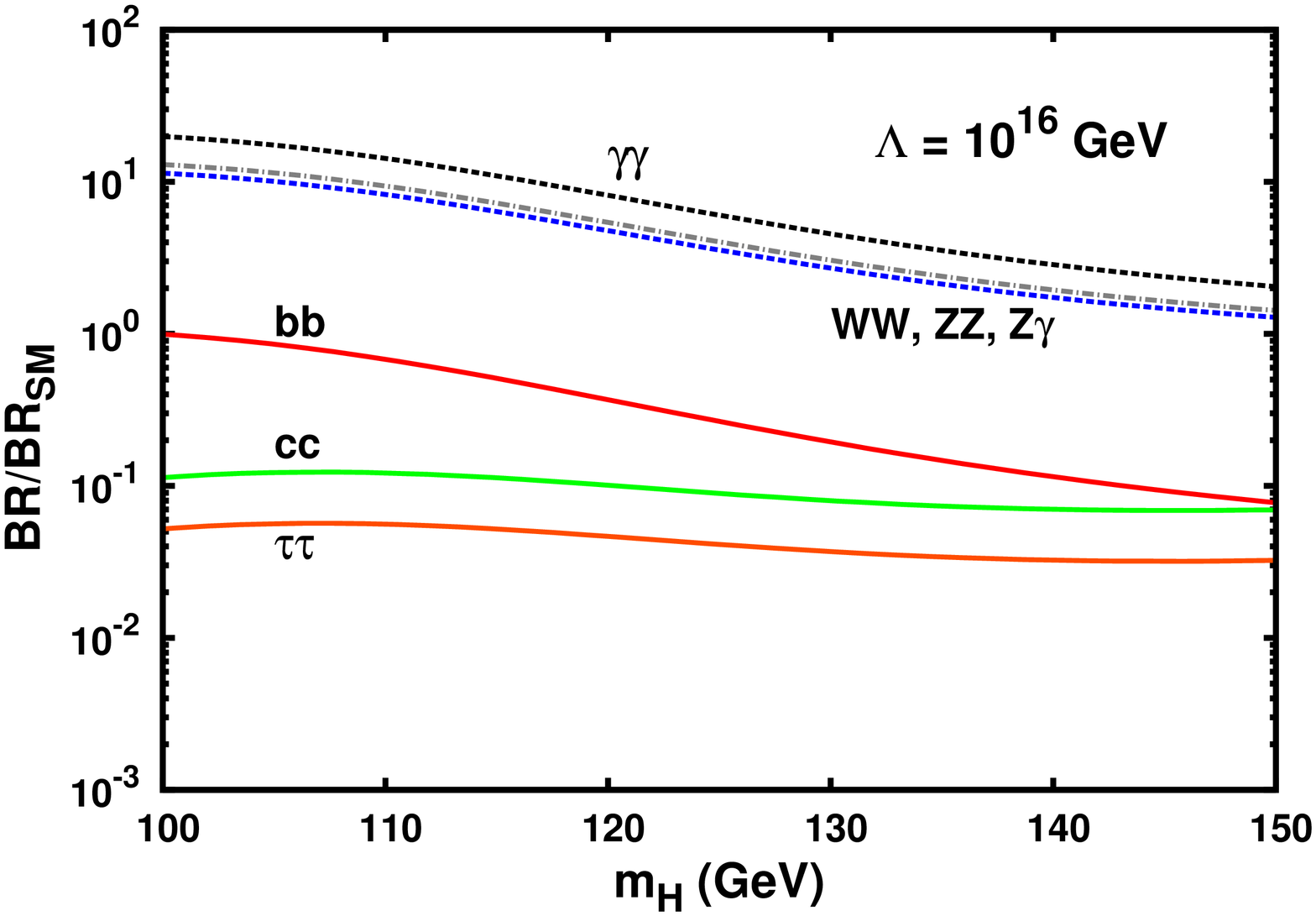, width=6cm,height=7.5cm, angle=-90}
\hspace{0cm}
\hfil
\epsfxsize=3.1in\epsfig{file=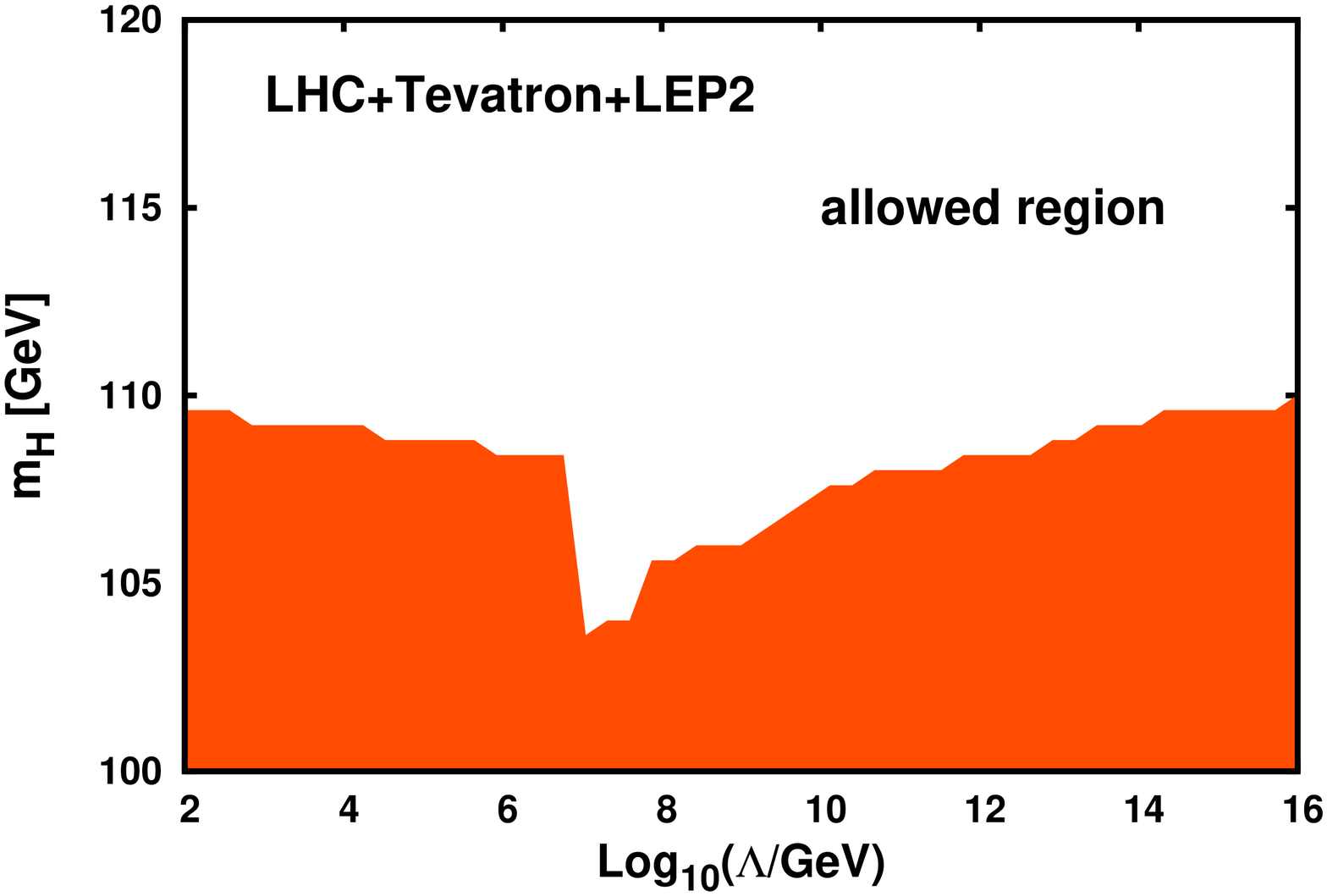,  width=6cm, height=7.5cm, angle=-90}}
\end{center}
\caption{\small Upper-left  plot shows 
 the total Higgs width 
$\Gamma_H$ versus $m_H$, for  
$\Lambda=10^{4,6,10,16}$ GeV.
Upper-right plot presents
 the Higgs BR for different decays, at $m_H=120$ GeV, 
versus the scale $\Lambda$. 
Lower-left plot 
 shows the Higgs BR' s normalized to their SM values versus $m_H$, for $\Lambda=10^{16}$ GeV. 
 Lower-right plot presents 
 the exclusion regions at 95\% C.L. (orange/grey area) 
in the ($m_H$, $\Lambda$) plane, as obtained through the code
Higgsbounds, version 3.4.0beta.}
\label{fig1}
\end{figure}
%%%%%%%%%%%%%%%%%%%%%%%%%%%%%%%%%%%%%%%%%%%%%%%%%%%%%%%%%%%%%%%%%%%%%%%%%%%
%%%%%%%%%%%%%%%%%%%%%%%%%%%%%%%  FIG 2 %%%%%%%%%%%%%%%%%%%%%%%%%%%%%%%%%%%%%%
\begin{figure}[tpb]
\begin{center}
\centerline{
\epsfxsize=3.1in\epsfig{file=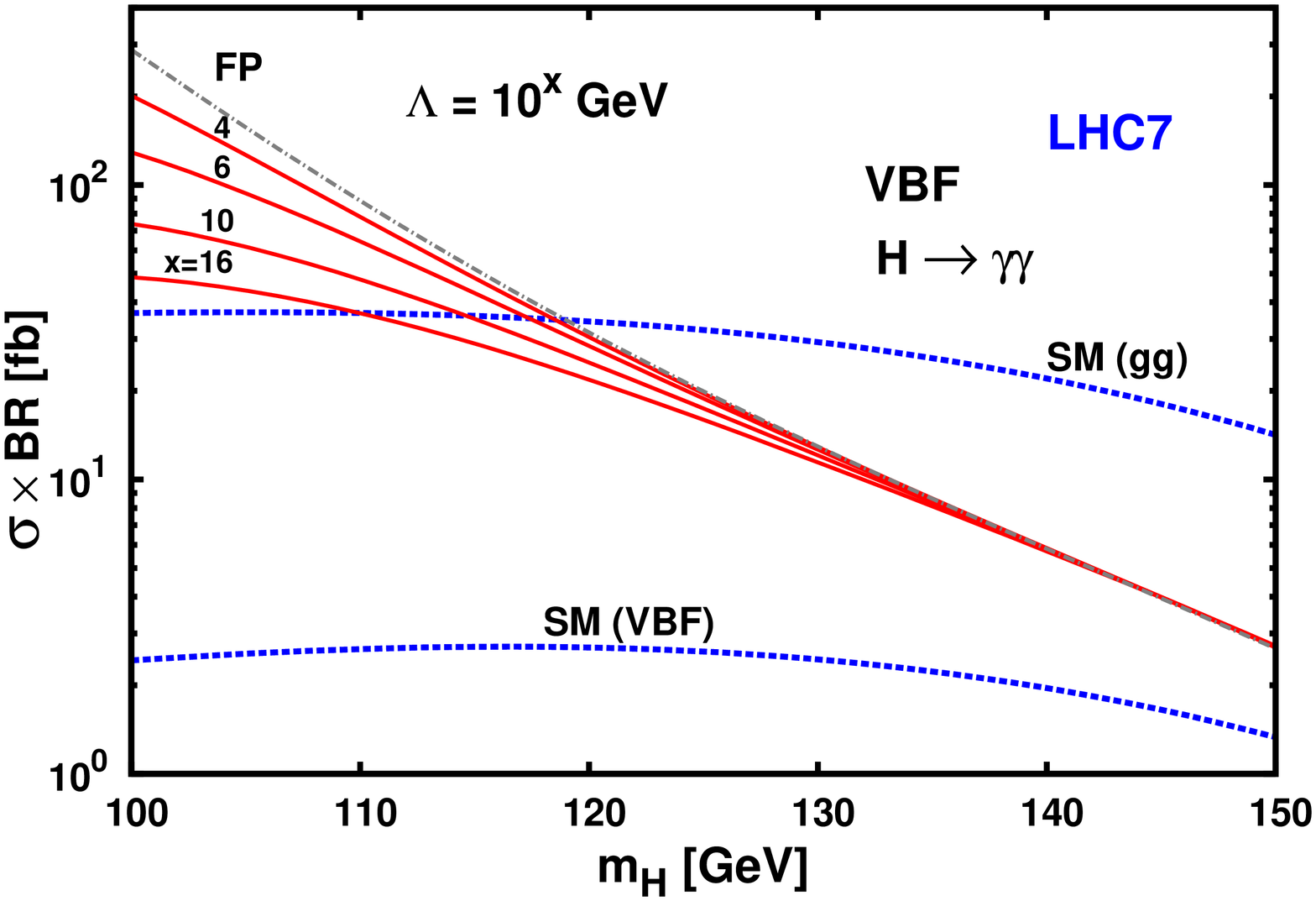, width=6cm,height=7.5cm, angle=-90}
\hspace{0cm}
\hfil
\epsfxsize=3.1in\epsfig{file=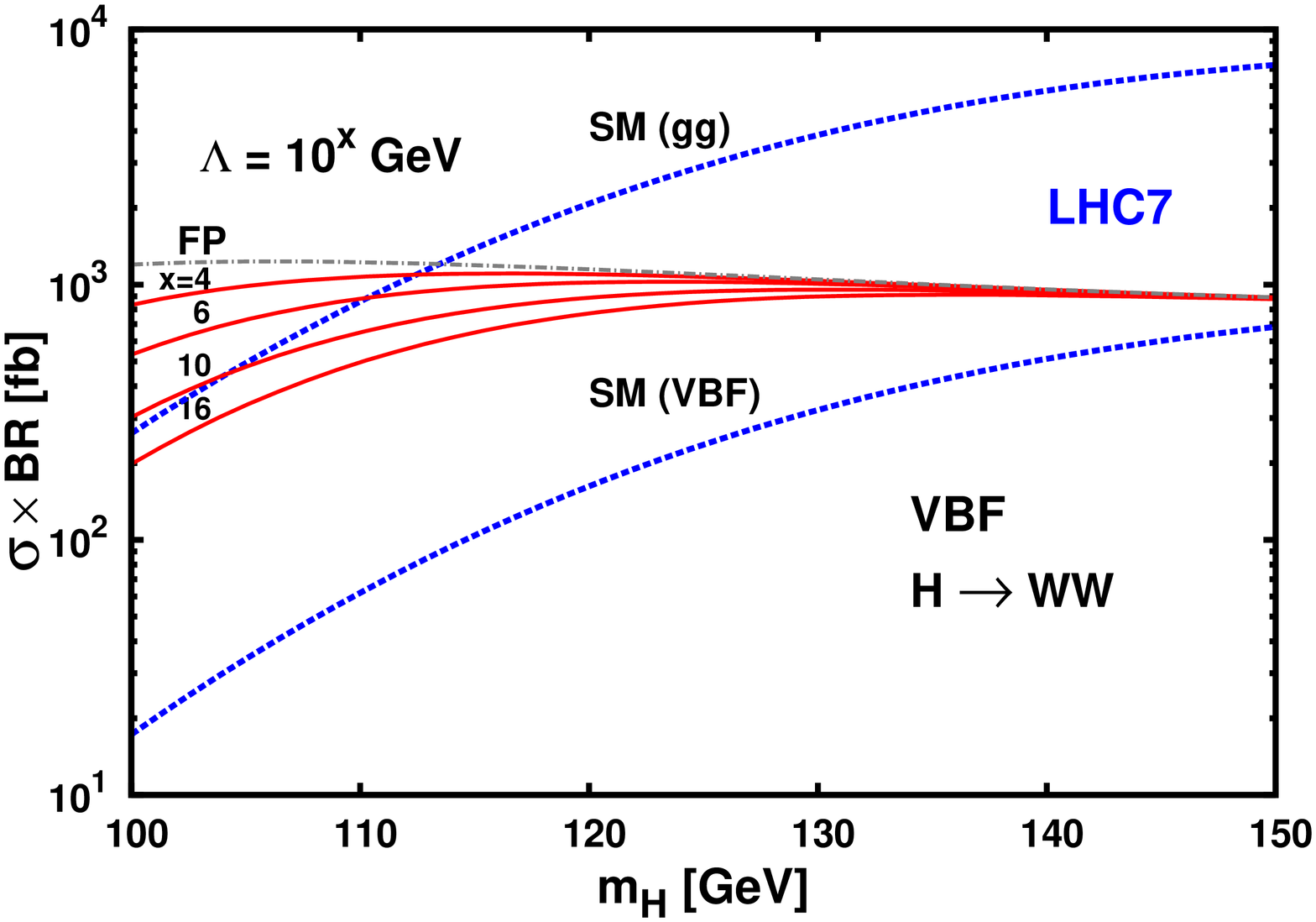,  
width=6cm, height=7.5cm, angle=-90}}

\vspace{0.5cm}
\centerline{
\epsfxsize=3.1in\epsfig{file=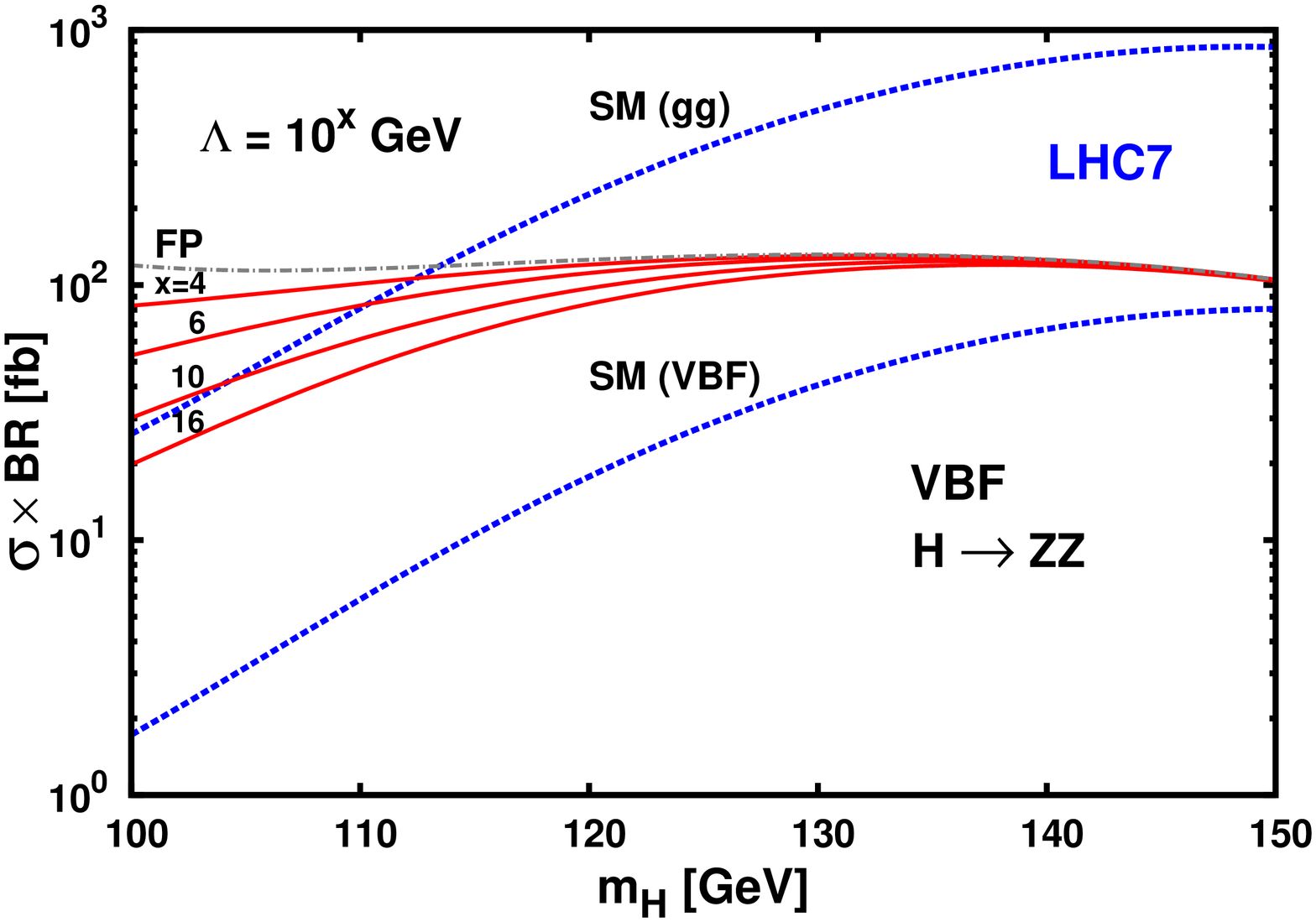, width=6cm,height=7.5cm, angle=-90}
\hspace{0cm}
\hfil
\epsfxsize=3.1in\epsfig{file=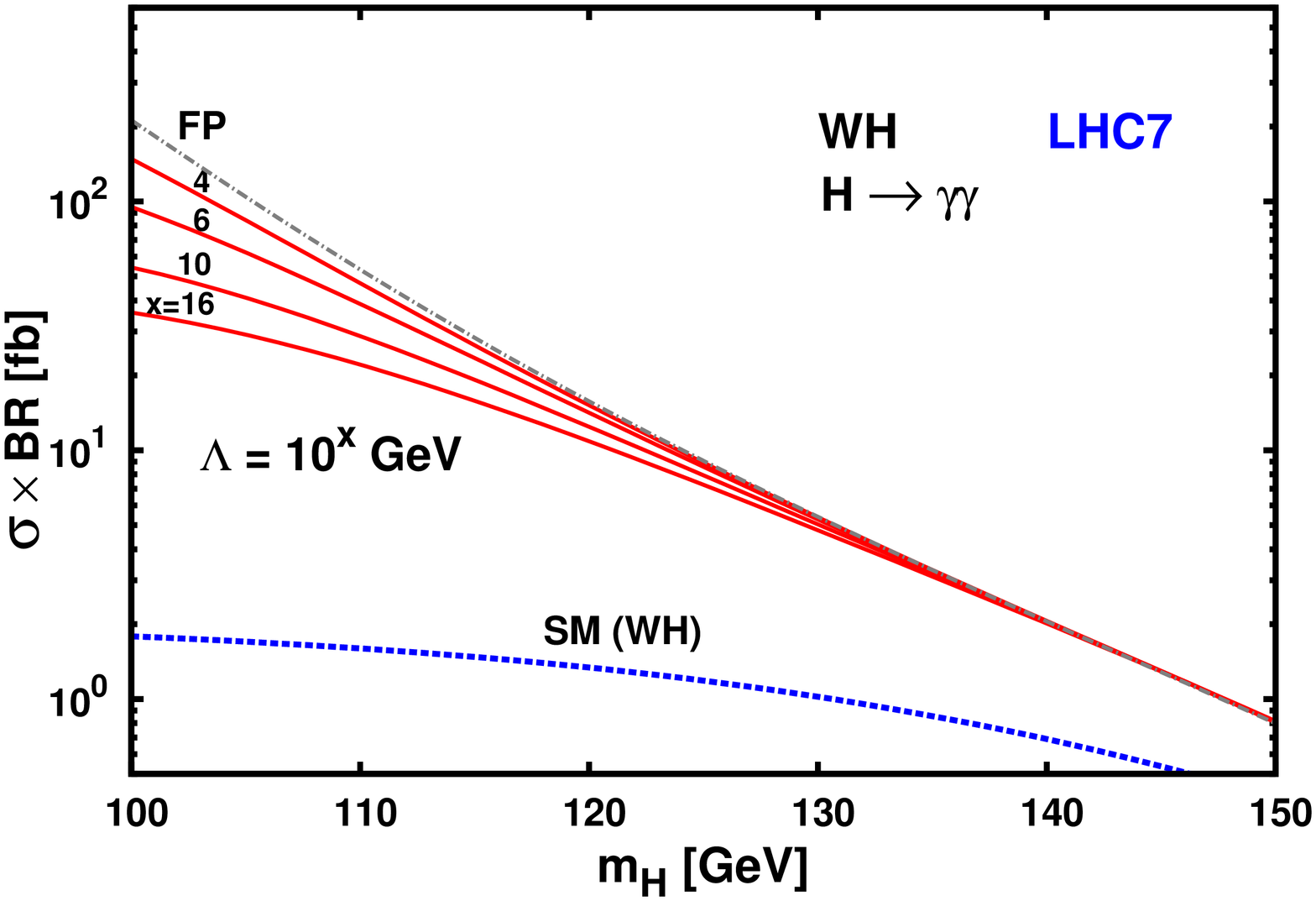,  width=6cm, height=7.5cm, angle=-90}}
\end{center}
\caption{\small Total 
cross sections times  Higgs branching ratio (BR) 
at the LHC at $\sqrt{S}=$ 7 TeV, for different Higgs production mechanism and decay channels, for a set of $\Lambda$ values. 
The upper-left, upper-right and lower-left plots
correspond to the Vector-Boson-Fusion (VBF) production and  decays  $H\to\gamma \gamma$
$H\to WW $, and $H\to Z Z$, respectively, while
the lower-right plot shows the rate for 
associated Higgs-W production (WH) and 
decay  $H\to \gamma \gamma $. SM curves are also shown for comparison.
}
\label{fig2}
\end{figure}
%%%%%%%%%%%%%%%%%%%%%%%%%%%%%%%%%%%%%%%%%%%%%%%%%%%%%%%%%%%%%%%%%%%%%%%%%%%
%%%%%%%%%%%%%%%%%%%%  FIG 3 %%%%%%%%%%%%%%%%%%%%%%%%%%%%%%%%%%%%%%
\begin{figure}[tpb]
\begin{center}
\centerline{
\epsfxsize=3.1in\epsfig{file=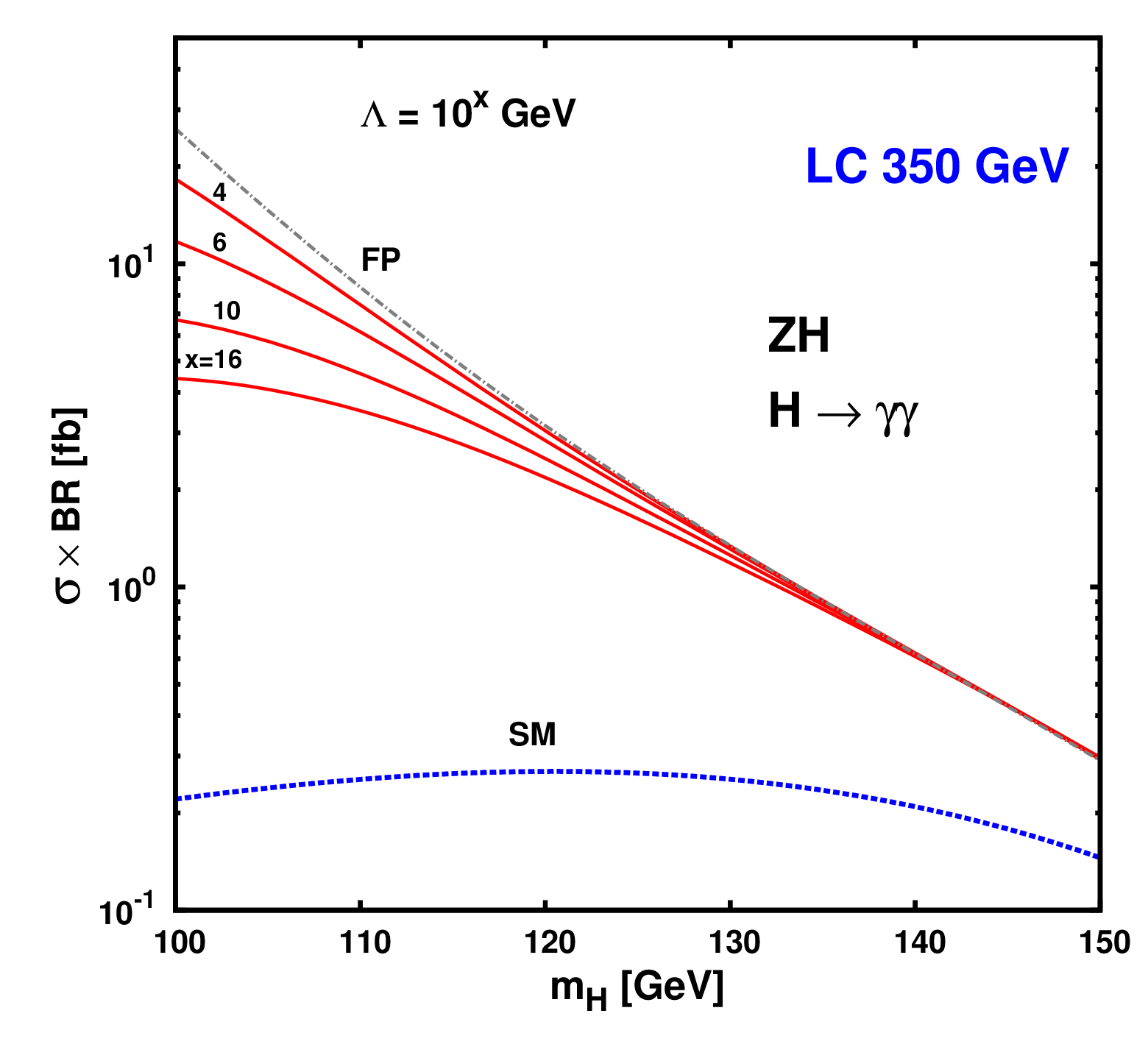, width=6cm,height=7.5cm, angle=-90}
\hspace{0cm}
\hfil
\epsfxsize=3.1in\epsfig{file=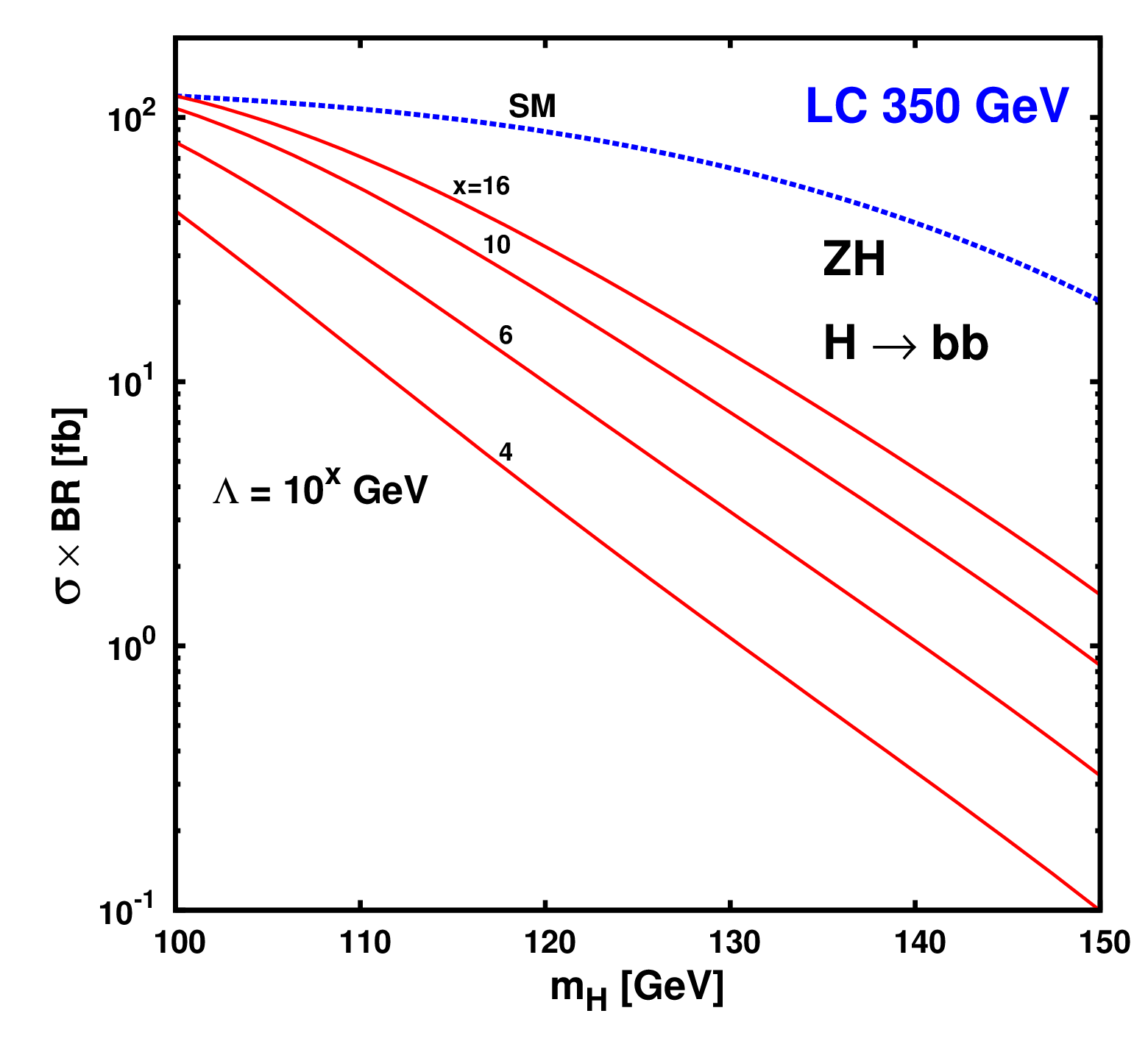,  
width=6cm, height=7.5cm, angle=-90}}

\vspace{0.5cm}
\centerline{
\epsfxsize=3.1in\epsfig{file=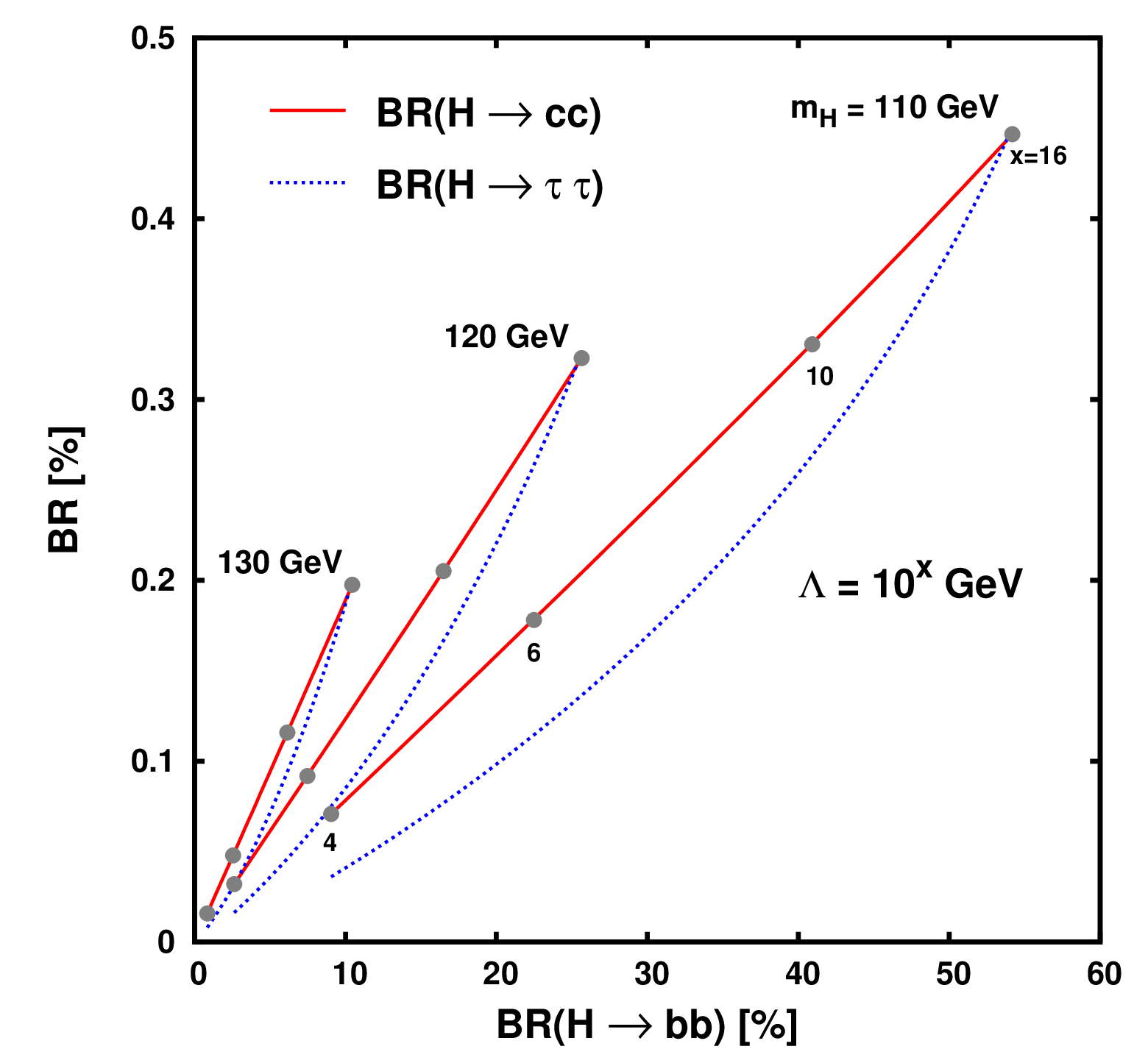, width=6cm,height=7.5cm, angle=-90}
\hspace{0cm}
\hfil
\epsfxsize=3.1in\epsfig{file=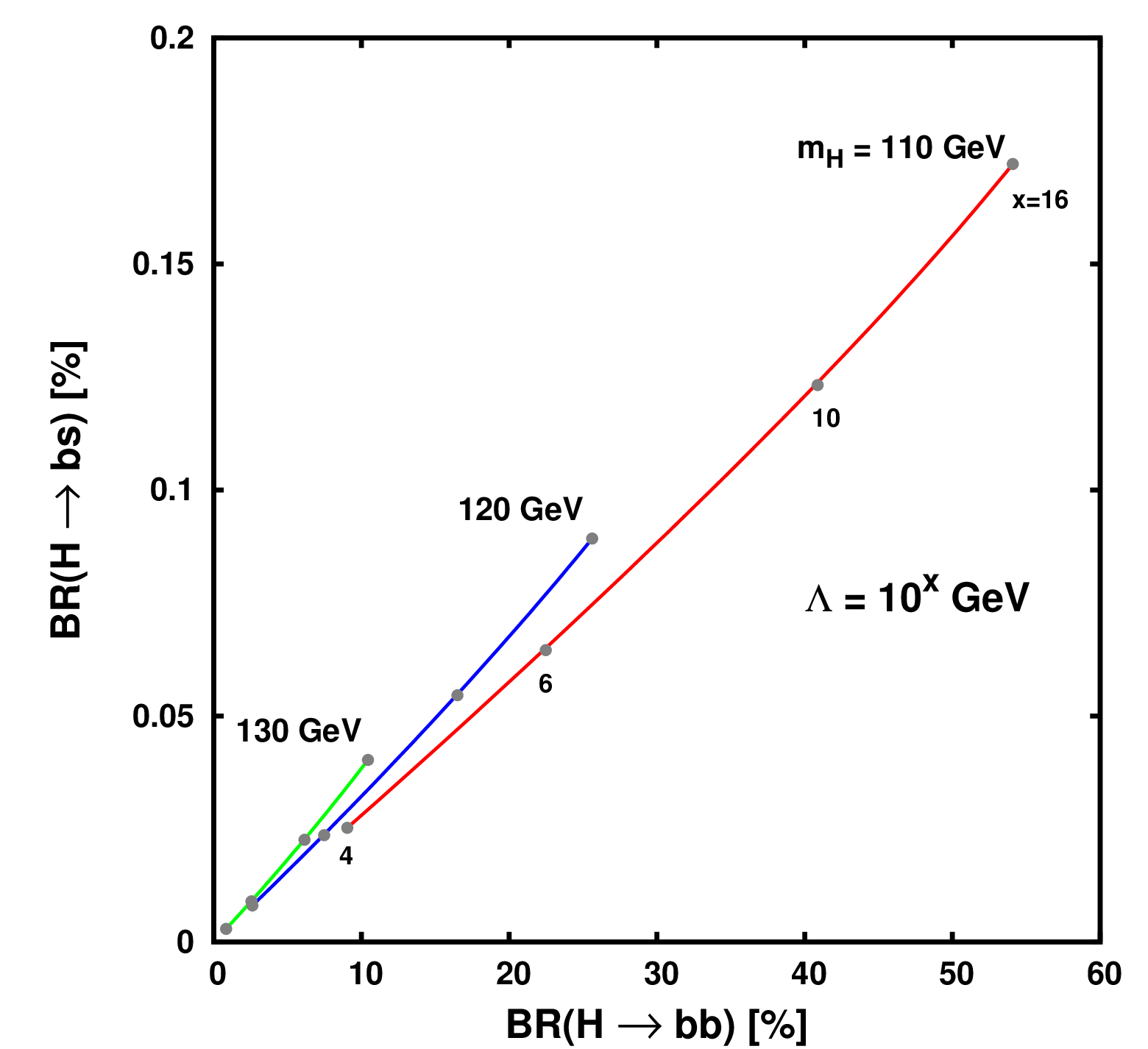,  width=6cm, height=7.5cm, angle=-90}}
\end{center}
\caption{\small 
Upper-left- and upper-right plots show to the total 
cross sections times the Higgs branching ratios 
BR for the decays $H\to \gamma \gamma$ and $H\to bb$, respectively,
at a linear collider with $\sqrt{S}=$ 350 GeV. 
SM curves are also shown for comparison.
Also shown are the correlations of the Higgs branching ratios 
[BR($H\to cc$), BR($H\to \tau \tau$)] (lower-left)
and the flavor-changing  
BR$(H\to bs)$  (lower-right), versus  BR($H\to bb)$, for different  values
of  $m_H=110,120,130$ GeV and  $\Lambda$.
}
\label{fig3}
\end{figure}
%%%%%%%%%%%%%%%%%%%%%%%%%%%%%%%%%%%%%%%%%%%%%%%%%%%%%%%%%%%%%%%%%%%%%%%%%%%%* LEP  
%?109.7 GeV (comb. ?? data on 4 exps) ('02) 
%                       (108.3 GeV, comb. ??+WW* in L3) 
%   * CDF  (100-) 
%? 114 GeV 7 fb-1 (May 2011) 
%   * D0    (100-) 
%? 112.9 GeV 8.2 fb-1 (July 2011) 
%   * CMS  (110-) 
%? 112 GeV 1.7 fb-1 (LP2011) 
%
%(new)Combining  H???, WW at Tevatron M<119 GeV@95%CL
%(Weiming Yao (LBNL) o A. Patwa (BNL)  HCP Parigi)
%
%110 -- 111 GeV and 113.5 -- 117.5 GeV.
%ATLAS-CONF-2011-149
%O.ÊIgonkinaÊ [110-?112]ÊandÊ[113-?118]ÊGeVÊmassÊrangeÊ@Ê95%CL
%%%%%%%%%%%%%%%%%%%%%%%%%%%%%%%%%%%%%%%%%%%%%%%%%%%%%%%%%%%%%%%%%%%%%%%%%%%%%
% ****************************************************************************
% BIBLIOGRAPHY AREA
%% ****************************************************************************
%\section*{Acknowledgments}
%
\begin{footnotesize}
% IF YOU DO NOT USE BIBTEX, USE THE FOLLOWING SAMPLE SCHEME FOR THE REFERENCES
% ----------------------------------------------------------------------------
%%%%%%%%%%%%%%%%%%%%%%%%%%%%%%%%%%%%%%%%%%%%%%%%%%%%%%%%%%%%%%%%%%%%%%%%%%%%%

\end{footnotesize}

% ****************************************************************************
% END OF BIBLIOGRAPHY AREA
% ****************************************************************************


\begin{thebibliography}{99}
\bibitem{djouadi}
%\cite{Djouadi:2005gi}
  A.~Djouadi,
  %``The Anatomy of electro-weak symmetry breaking. I: The Higgs boson in the
  %standard model,''
  Phys.\ Rept.\  {\bf 457} (2008) 1
  [arXiv:hep-ph/0503172].
  %%CITATION = PRPLC,457,1;%%
%======================================================================
\bibitem{LHCHiggsCrossSectionWG}
LHC Higgs Cross Section Working Group, webpage \\
https://twiki.cern.ch/twiki/bin/view/LHCPhysics/Fermiophobic
%\bibitem{LHCHiggsCrossSectionWorkingGroup:2011ti}
%  LHC Higgs Cross Section Working Group, S.~Dittmaier, C.~Mariotti, G.~Passarino, and R.~Tanaka (Eds.), 
%  {\sl Handbook of LHC Higgs Cross Sections: 1. Inclusive Observables}, 
%  CERN-2011-002 (CERN, Geneva, 2011), {\tt arXiv:1101.0593 [hep-ph]} %======================================================================
\bibitem{noi-uno}
%\cite{Gabrielli:2010cw}
  E.~Gabrielli and B.~Mele,
  %``Testing Effective Yukawa Couplings in Higgs Searches at the Tevatron and
  %LHC,''
  Phys.\ Rev.\  D {\bf 82}, 113014 (2010)
  [Erratum-ibid.\  D {\bf 83}, 079901 (2011)]
  [arXiv:1005.2498 [hep-ph]].
  %%CITATION = PHRVA,D82,113014;%%
%======================================================================
\bibitem{HiggsFP}
%\cite{Gunion:1989ci}
  J.~F.~Gunion, R.~Vega and J.~Wudka,
  %``Higgs triplets in the standard model,''
  Phys.\ Rev.\  D {\bf 42} (1990) 1673;
  %%CITATION = PHRVA,D42,1673;%%
%\cite{Bamert:1993ah}
  P.~Bamert and Z.~Kunszt,
  %``Gauge boson masses dominantly generated by Higgs triplet contributions?,''
  Phys.\ Lett.\  B {\bf 306} (1999) 335
  [arXiv:hep-ph/9303239];
  %%CITATION = PHLTA,B306,335;%%
%\cite{Akeroyd:1995hg}
  A.~G.~Akeroyd,
  %``Fermiophobic Higgs bosons at the Tevatron,''
  Phys.\ Lett.\  B {\bf 368} (1996) 89
  [arXiv:hep-ph/9511347];
  %%CITATION = PHLTA,B368,89;%%
%\cite{Barroso:1999bf}
  A.~Barroso, L.~Brucher and R.~Santos,
  %``Is there a light fermiophobic Higgs?,''
  Phys.\ Rev.\  D {\bf 60} (1999) 035005
  [arXiv:hep-ph/9901293];
  %%CITATION = PHRVA,D60,035005;%%
%\cite{Brucher:1999tx}
  L.~Brucher and R.~Santos,
  %``Experimental signatures of fermiophobic Higgs bosons,''
  Eur.\ Phys.\ J.\  C {\bf 12} (2000) 87
  [arXiv:hep-ph/9907434].
  %%CITATION = EPHJA,C12,87;%%
%======================================================================
\bibitem{arason}
%\cite{Arason:1991ic}
  H.~Arason, D.~J.~Castano, B.~Keszthelyi, S.~Mikaelian, E.~J.~Piard, P.~Ramond and B.~D.~Wright,
  %``Renormalization group study of the standard model and its extensions. 1.
  %The Standard model,''
  Phys.\ Rev.\  D {\bf 46} (1992) 3945.
  %%CITATION = PHRVA,D46,3945;%% %======================================================================
\bibitem{noi-due}
%\cite{Gabrielli:2011yn}
  E.~Gabrielli and B.~Mele,
  %``Effective Yukawa couplings and flavor-changing Higgs boson decays at linear
  %colliders,''
  Phys.\ Rev.\  D {\bf 83}, 073009 (2011)
  [arXiv:1102.3361 [hep-ph]].
  %%CITATION = PHRVA,D83,073009;%%
%======================================================================
%======================================================================
\bibitem{FP-LEP}
%\cite{Heister:2002ub}
  A.~Heister {\it et al.}  [ALEPH Collaboration],
  %``Search for gamma gamma decays of a Higgs boson in e+ e- collisions at
  %s**(1/2) up to 209-GeV,''
  Phys.\ Lett.\  B {\bf 544} (2002) 16;
  %%CITATION = PHLTA,B544,16;%%
%\cite{Abreu:2001ib}
  P.~Abreu {\it et al.}  [DELPHI Collaboration],
  %``Search for a fermiophobic Higgs at LEP 2,''
  Phys.\ Lett.\  B {\bf 507} (2001) 89
  [arXiv:hep-ex/0104025];
  %%CITATION = PHLTA,B507,89;%%
%\cite{Achard:2002jh}
  P.~Achard {\it et al.}  [L3 Collaboration],
  %``Search for a Higgs boson decaying into two photons at LEP,''
  Phys.\ Lett.\  B {\bf 534} (2002) 28
  [arXiv:hep-ex/0203016];
  %%CITATION = PHLTA,B534,28;%%
%\cite{Achard:2003jb}
 % P.~Achard {\it et al.}  [L3 Collaboration],
  %``Search for a Higgs boson decaying to weak boson pairs at LEP,''
  Phys.\ Lett.\  B {\bf 568}, (2003) 191
  [arXiv:hep-ex/0307010];
  %%CITATION = PHLTA,B568,191;%%
  G.~Abbiendi {\it et al.}  [OPAL Collaboration],
  %``Search for associated production of massive states decaying into two
  %photons in $e^{+} e^{-}$ annihilations at $\sqrt{s}$ = 88-GeV to 209-GeV,''
  Phys.\ Lett.\  B {\bf 544} (2002) 44
  [arXiv:hep-ex/0207027].
%======================================================================
 \bibitem{FP-Tevatron} 
 [CDF and D0 and Tevatron New Higgs Working Group Collaborations],
  ``Combined CDF and D0 Upper Limits on Fermiophobic Higgs Boson Production with up to 8.2 fb$^{-1}$ of $p\bar{p}$ data'',
  arXiv:1109.0576 [hep-ex].
%======================================================================
\bibitem{FP-CMS}
[CMS Collaboration], 
``Search for a Higgs boson decaying into two photons in the 
CMS detector", 
CMS PAS HIG-11-021.
%======================================================================
%======================================================================
%======================================================================\cite{SM-LEP,SM-Tevatron,SM-ATLAS,SM-CMS}.
%======================================================================
\bibitem{SM-LEP}
%\cite{Barate:2003sz}
  R.~Barate {\it et al.}  (LEP Working Group for Higgs boson searches and ALEPH, 
DELPHI, L3, and OPAL Collaborations),
  %``Search for the standard model Higgs boson at LEP,''
  Phys.\ Lett.\  B {\bf 565} (2003) 61
  [arXiv:hep-ex/0306033].
  %%CITATION = PHLTA,B565,61;%%
%======================================================================
\bibitem{SM-Tevatron}
  [TEVNPH (Tevatron New Phenomena and Higgs Working Group) and CDF and D0 Collaborations],
  ``Combined CDF and D0 Upper Limits on Standard Model Higgs Boson Production with up to 8.6 fb-1 of Data'',
  arXiv:1107.5518 [hep-ex].
%======================================================================
\bibitem{SM-ATLAS}
[ATLAS Collaboration], 
``Update of the Combination of Higgs Boson Searches in pp Collisions at 
$\sqrt(S)$ = 7 TeV with the ATLAS Experiment at the LHC",  
ATLAS-CONF-2011-135.
%======================================================================
\bibitem{SM-CMS}
[CMS Collaboration], 
``Combination of Higgs Searches", 
CMS PAS HIG-11-021.
%======================================================================
%\bibitem{HboundsEW}
%LEP Electroweak working group, http://www.cern.ch/LEPEWWG/ .
%%\cite{Amsler:2008zzb}
%======================================================================
\bibitem{Higgsbounds}
  P.~Bechtle, O.~Brein, S.~Heinemeyer, G.~Weiglein and K.~E.~Williams,
  %``HiggsBounds: Confronting Arbitrary Higgs Sectors with Exclusion Bounds from LEP and the Tevatron,''
  Comput.\ Phys.\ Commun.\  {\bf 181} (2010) 138
  [arXiv:0811.4169 [hep-ph]];
 %``HiggsBounds 2.0.0: Confronting Neutral and Charged Higgs Sector Predictions with Exclusion Bounds from LEP and the Tevatron,''
  Comput.\ Phys.\ Commun.\  {\bf 182} (2011) 2605
  [arXiv:1102.1898 [hep-ph]].
%======================================================================
\bibitem{ILC}
J.~A.~Aguilar-Saavedra {\it et al.}  [ECFA/DESY LC Physics Working Group],
  arXiv:hep-ph/0106315, 
   http://tesla.desy.de/new$\_$pages/TDR$\_$CD/start.html  .
%======================================================================

\end{thebibliography}
\end{document}